\documentclass[11pt,eqs]{article}
\usepackage{latexsym,amsmath}

\def\cleq{\setcounter{equation}{0}}
\title{Courant bracket as T-dual invariant extension of Lie bracket 
\thanks{Work supported in part by
the Serbian Ministry of Education and Science and Technological Development, under contract No. 171031.}}
\author{ Lj. Davidovi\'c \thanks{e-mail: ljubica@ipb.ac.rs}, I. Ivani\v sevi\'c \thanks{e-mail: ivanisevic@ipb.ac.rs} and B. Sazdovi\'c
\thanks{e-mail: sazdovic@ipb.ac.rs}\\
{\it Institute of Physics, University of Belgrade}\\
{\it Pregrevica 118, 11080 Belgrade, Serbia}}

\begin{document}
\maketitle
\begin{abstract}

We consider the symmetries of a closed bosonic string, starting with the general coordinate transformations. Their generator takes vector components $\xi^\mu$ as its parameter and its Poisson bracket algebra gives rise to the Lie bracket of its parameters. We are going to extend this generator in order for it to be invariant upon self T-duality, i.e. T-duality realized in the same phase space. The new generator is a function of a $2D$ double symmetry parameter $\Lambda$, that is a direct sum of vector components $\xi^\mu$, and 1-form components $\lambda_\mu$. The Poisson bracket algebra of a new generator produces the Courant bracket in a same way that the algebra of the general coordinate transformations produces Lie bracket. In that sense, the Courant bracket is T-dual invariant extension of the Lie bracket. When the Kalb-Ramond field is introduced to the model, the generator governing both general coordinate and local gauge symmetries is constructed. It is no longer self T-dual and its algebra gives rise to the $B$-twisted Courant bracket, while in its self T-dual description, the relevant bracket becomes the $\theta$-twisted Courant bracket. Next, we consider the T-duality and the symmetry parameters that depend on both the initial coordinates $x^\mu$ and T-dual coordinates $y_\mu$. The generator of these transformations is defined as an inner product in a double space and its algebra gives rise to the C-bracket. 
\end{abstract}

\section{Introduction}
\cleq
The Courant bracket \cite{courant,courant1} and various generalizations obtained by its twisting had been relevant to the string theory since its appearance in the algebra of generalized currents \cite{c,nick1,nick2,crdual}. It represents the generalization of the Lie bracket on spaces of generalized vectors, understood as the direct sum of the elements of the tangent bundle and the elements of the cotangent bundle. Although the Lie bracket satisfies the Jacobi identity, the Courant bracket does not. Its Jacobiator is equal to the exterior derivative of the Nijenhuis operator. 

It is well known that the commutator of two general coordinate transformations along two vector fields produces another general coordinate transformation along the vector field equal to their Lie bracket. Since the Courant bracket represents its generalization, it is worth considering how it is related to symmetries of the bosonic string $\sigma$-model. 

In \cite{doucou}, the field theory defined on the double torus, and its symmetries for restricted parameters were considered. The double space is seen as a direct sum of the initial and T-dual phase space, and the background fields depend on both of these coordinates.  The symmetry algebra is closed only for restricted parameters, defined on the same isotropic space, in which case it gives rise to the C-bracket as the T-dual invariant bracket. The C-bracket \cite{siegel1,siegel} is the bracket that generalizes the Lie bracket on double space. 

In this paper, we analyze the general classical bosonic string $\sigma$-model and algebra of its symmetries generators, where both the background fields and symmetry parameters depend only on the coordinates $x^\mu$. We firstly consider the closed bosonic string moving in the background characterized solely by the metric tensor. We extend the generator of the general coordinate transformations so that it becomes invariant upon self T-duality, understood as T-duality realized in the same phase space \cite{crdual}. We obtain the Courant bracket in the Poisson bracket algebra of this extended generator. The Courant bracket is therefore a self T-dual invariant extension of the Lie bracket. 

Furthermore, we consider the bosonic string $\sigma$-model that includes the antisymmetric Kalb-Ramond field too. The antisymmetric field is introduced by the action of B-transformation on the generalized metric. We construct the symmetry generator and recognize that it generates both the general coordinate and the local gauge transformations \cite{dualsim}. In this case, the symmetry generator is not invariant upon self T-duality and it gives rise to the twisted Courant bracket. The matrix that governs this twist is exactly the matrix of B-shifts.

Next, we consider the self T-dual description of the theory, that we construct in the analogous manner, this time with the action of $\theta$-transformation, T-dual to the B-transformation. We obtain the bracket governing the generator algebra that turns out to be the $\theta$-twisted Courant bracket, also known as the Roytenberg bracket \cite{nick1,royt}. The twisted Courant and Roytenberg brackets had been shown to be related by self T-duality \cite{crdual}.

Lastly, we consider the more conventional T-duality, connecting different phase spaces. We generalize our results, by demanding that the symmetry parameters depend on both the initial and T-dual coordinates. We consider the symmetry generator that is a sum of the generator of general coordinate transformations and its analogous generator in the T-dual phase space. In this case, additional constraints, similar to the ones in \cite{doucou, siegel1,siegel}, have to be imposed on symmetry parameters, in order for the generator algebra to be closed. We extend the Poisson bracket relations for both initial and T-dual phase spaces and obtain the generator algebra, which produces the $C$-bracket. The $C$ bracket is the generalization of the Courant bracket when parameters depend on both initial and T-dual coordinates. The invariance upon T-duality is guaranteed from the way how the bracket is obtained. If parameters do not depend on T-dual coordinates, $C$-bracket reduces to the Courant bracket.

\section{Bosonic string moving in the background characterized by the metric field}
\cleq

Consider the closed bosonic string, moving in the background defined by the coordinate dependent metric field $G_{\mu \nu}(x)$, with the Kalb-Ramond field set to zero $B_{\mu \nu} = 0$ and the constant dilaton field $\Phi = const$. In the conformal gauge, the Lagrangian density is given by \cite{action, regal}
\begin{equation}\label{eq:action}
{\cal{L}} = \frac{\kappa }{2}\eta^{\alpha\beta}G_{\mu\nu}(x) \partial_{\alpha}x^{\mu}\partial_{\beta}x^{\nu} \, ,
\end{equation}
where $x^\mu (\xi),\ \mu = 0, 1, ..., D-1$ are coordinates on the $D$-dimensional space-time, and $\eta^{\alpha \beta}, \ \alpha, \beta = 0,1$ is the worldsheet metric, $\epsilon^{01} = -1$ is the Levi-Civita symbol, and $\kappa = \frac{1}{2 \pi \alpha^\prime}$ with $\alpha^\prime$ being the Regge slope parameter. The Legendre transformation of the Lagrangian gives the canonical Hamiltonian
\begin{equation} \label{eq:HG}
{\cal H}_{C} = \pi_\mu \dot{x}^\mu - \mathcal{L} = \frac{1}{2 \kappa} \pi_\mu (G^{-1})^{\mu \nu} \pi_\nu + \frac{\kappa}{2} x^{\prime \mu} G_{\mu \nu} x^{\prime \nu} \, ,
\end{equation}
where $\pi_\mu$ are canonical momenta conjugate to coordinates $x^\mu$, given by
\begin{equation} \label{eq:pidef}
\pi_\mu =
\frac{\partial{\mathcal{L}}}{\partial{\dot{x}^{\mu}}} =
\kappa G_{\mu\nu}(x) \dot{x}^{\nu} \, .
\end{equation}
The Hamiltonian can be rewritten in the matrix notation
\begin{equation} \label{eq:Hcmat}
{\cal H}_{C} = \frac{1}{2\kappa} (X^T)^M G_{MN} X^N \, , 
\end{equation}
where $X^M$ is a double canonical variable, given by 
\begin{equation} \label{eq:Xdouble}
X^M = \begin{pmatrix}
\kappa x^{\prime \mu} \\
\pi_\mu \\
\end{pmatrix}\, ,
\end{equation}
and $G_{MN}$ is the so called generalized metric, that in the absence of the Kalb-Ramond field takes the diagonal form
\begin{equation} \label{eq:GMN}
G_{MN} = 
\begin{pmatrix}
G_{\mu \nu} & 0 \\
0 & (G^{-1})^{\mu \nu}
\end{pmatrix} \, .
\end{equation}

In this paper, we firstly consider the T-duality realized without changing the phase space, which is called the self T-duality \cite{crdual}. Two quantities are said to be self T-dual if they are invariant upon 
\begin{equation} \label{eq:xpidual}
\kappa x^{\prime \mu} \leftrightarrow \pi_\mu \, , \ \ G_{\mu \nu} \leftrightarrow {^\star G}^{\mu\nu} =
(G^{-1})^{\mu\nu} \, .
\end{equation}
The first part of (\ref{eq:xpidual}) corresponds to the T-duality interchanging the winding and momentum numbers, which are respectively obtained by integrating $\kappa x^{\prime \mu}$ and $\pi_\mu$   over the worldsheet space parameter $\sigma$ \cite{tdwcb}. The second part of (\ref{eq:xpidual}) corresponds to swapping the background fields for the T-dual background fields. Our approach gives the same expression for the T-dual metric as the usual T-dualization procedure obtained by Buscher in the special case of zero Kalb-Ramond field \cite{buscher,tdual,tdual1}.

\subsection{Symmetry generator}
Let us consider symmetries of the closed bosonic string. The canonical momenta $\pi_\mu$ generate the general coordinate transformations. The generator is given by \cite{dualsim}
\begin{equation} \label{eq:GCTdef}
{\cal G}_{GCT} (\xi) =\int_0^{2\pi} d\sigma \xi^\mu (x) \pi_\mu \, ,
\end{equation}
with $\xi^\mu$ being a symmetry parameter. The general coordinate transformations of the metric tensor are given by \cite{doucou,dualsim}
\begin{equation} \label{eq:GGCT}
\delta_\xi G_{\mu \nu} = {\cal L}_\xi G_{\mu \nu} \, ,
\end{equation}
where ${\cal L}_\xi $ is the Lie derivative along the vector field $\xi$. Its action on the metric field is
\begin{eqnarray}\label{eq:LieG}
{\cal L}_\xi G_{\mu \nu} = D_\mu \xi_\nu + D_\nu \xi_\mu \, ,
\end{eqnarray}
where $D_\mu$ are covariant derivatives defined in a usual way 
\begin{equation}
D_\mu \xi_\nu = \partial_\mu \xi_\nu - \Gamma^{\rho}_{\mu \nu} \xi_\rho \, ,
\end{equation}
and
$\Gamma^\mu_{\nu\rho}
=\frac{1}{2}(G^{-1})^{\mu\sigma}(\partial_\nu G_{\rho\sigma}+\partial_\rho G_{\sigma\nu}
-\partial_\sigma G_{\nu\rho})$ are Christoffel symbols. It is easy to verify, using the standard Poisson bracket relations
\begin{equation} \label{eq:PBR}
\{ x^{\mu} (\sigma), \pi_\nu (\bar{\sigma}) \} = \delta^\mu_{\ \nu} \delta(\sigma - \bar{\sigma}) \, ,
\end{equation}
that the Poisson bracket of these generators can be written as
\begin{eqnarray}\label{eq:GCTalg}
\Big\{ {\cal G}_{GCT}(\xi_1), \, {\cal G}_{GCT}(\xi_2) \Big\} = -{\cal G}_{GCT}\Big( [\xi_1, \xi_2 ]_L\Big) \, ,
\end{eqnarray}
where $[\xi_1, \xi_2 ]_L$ is the Lie bracket. The Lie bracket is the commutator of two Lie derivatives
\begin{eqnarray}
[\xi_1, \xi_2 ]_L = {\cal L}_{\xi_1} {\xi_2} - {\cal L}_{\xi_2} {\xi_1} \equiv {\cal L}_{\xi_3} \, ,
\end{eqnarray}
which results in another Lie derivative along the vector $\xi_3^\mu$, given by
\begin{eqnarray}\label{eq:liebr}
\xi_3^\mu = \xi_1^\nu \partial_\nu \xi_2^\mu - \xi_2^\nu \partial_\nu \xi_1^\mu \, . 
\end{eqnarray}

Let us now construct the symmetry generator that is related to the generator of general coordinate transformations by self T-duality (\ref{eq:xpidual}) 
\begin{equation}\label{eq:Tdual}
{\cal G}_{LG} (\lambda) =\int_0^{2\pi} d\sigma \lambda_\mu(x) \kappa x^{\prime\mu} \, ,
\end{equation}
where $\lambda_\mu$ is a gauge parameter. 

The symmetry parameters $\xi^\mu$ and $\lambda_\mu$ are vector and 1-form components, respectively. They can be combined in a double gauge parameter, given by 
\begin{equation} \label{eq:Lxi}
\Lambda^M = \begin{pmatrix}
\xi^\mu \\
\lambda_\mu \\
\end{pmatrix} \, .
\end{equation}
The double gauge parameter is a generalized vector, defined on the direct sum of elements of tangent and cotangent bundle. 
Combining (\ref{eq:GCTdef}) and (\ref{eq:Tdual}), we obtain the symmetry generator that is self T-dual (\ref{eq:xpidual})
\begin{equation}\label{eq:gltilde}
{\cal G} (\xi, \lambda)= {\cal G}_{GCT} (\xi) + {\cal G}_{LG} (\lambda) =
\int_0^{2\pi} d\sigma\Big[\xi^\mu\pi_\mu+ \lambda_\mu \kappa x^{\prime\mu} \Big] = \int_0^{2\pi} d\sigma (\Lambda^T)^M \eta_{MN} X^N \, ,
\end{equation}
where $\eta_{MN}$ is the $O(D,D)$ invariant metric \cite{ODD}, given by 
\begin{equation} \label{eq:Omega}
\eta_{MN} = 
\begin{pmatrix}
0 & 1 \\
1 & 0 
\end{pmatrix} \, .
\end{equation}
The expression $ (\Lambda^T)^M \eta_{MN} X^N$ can be recognized as the natural inner product on the space of generalized vectors
\begin{equation} \label{eq:skalproizvod}
\langle \Lambda^M, X^N \rangle = (\Lambda^T)^M \eta_{MN} X^N \, .
\end{equation}

We are interested in obtaining the algebra of this extended symmetry generator (\ref{eq:gltilde}), analogous to (\ref{eq:GCTalg}). Using the Poisson bracket relations (\ref{eq:PBR}), we obtain 
\begin{align} \label{eq:Gpom}
\Big\{ {\cal G} (\xi_1, \lambda_1), \, {\cal G} (\xi_2, \lambda_2) \Big\} = & \int d\sigma \Big[ \pi_\mu (\xi_2^\nu \partial_\nu \xi_1^\mu - \xi_1^\nu \partial_\nu \xi_2^\mu) + \kappa x^{\prime \mu} (\xi_2^\nu \partial_\nu \lambda_{1 \mu} - \xi_1^\nu \partial_\nu \lambda_{2 \mu}) \Big] \\ \notag
& + \int d\sigma d\bar{\sigma} \kappa \Big[\lambda_{1 \mu} (\sigma) \xi_2^\mu (\bar{\sigma}) + \lambda_{2 \mu} (\bar{\sigma}) \xi^\mu_1 (\sigma) \Big] \partial_\sigma \delta(\sigma-\bar{\sigma}) \, .
\end{align}
In order to transform the anomalous part, we note that
\begin{equation} \label{eq:deltapola}
\partial_\sigma \delta(\sigma-\bar{\sigma}) = \frac{1}{2} \partial_\sigma \delta(\sigma-\bar{\sigma}) -\frac{1}{2} \partial_{\bar{\sigma}} \delta(\sigma-\bar{\sigma}) \, ,
\end{equation}
and
\begin{equation} \label{eq:fdelta}
f(\bar{\sigma}) \partial_\sigma \delta(\sigma-\bar{\sigma}) = f(\sigma) \partial_\sigma \delta(\sigma-\bar{\sigma})+f^\prime (\sigma) \delta(\sigma-\bar{\sigma}) \, .
\end{equation}
Applying the previous two relations to the right hand side of (\ref{eq:Gpom}), one obtains
\begin{equation} \label{eq:GGG}
\Big\{ {\cal G} (\xi_1, \lambda_1), \, {\cal G} (\xi_2, \lambda_2) \Big\} = -{\cal G} (\xi, \lambda) \, ,
\end{equation}
where the resulting gauge parameters are given by
\begin{eqnarray} \label{eq:courantt}
\xi^\mu &=& \xi_1^\nu \partial_\nu \xi_2^\mu - \xi_2^\nu \partial_\nu \xi_1^\mu \, , \\ \notag
\lambda_\mu &=& \xi_1^\nu (\partial_\nu \lambda_{2 \mu} - \partial_\mu \lambda_{2 \nu}) - \xi_2^\nu (\partial_\nu \lambda_{1 \mu} - \partial_\mu \lambda_{1 \nu})+\frac{1}{2} \partial_\mu (\xi_1 \lambda_2- \xi_2 \lambda_1 ) \, .
\end{eqnarray}
These relations define the Courant bracket $[(\xi_1,\lambda_1), (\xi_2,\lambda_2)]_C = (\xi,\lambda)$ \cite{courant, courant1}, allowing us to rewrite the generator algebra (\ref{eq:GGG}) 
\begin{equation} \label{eq:GGcourant}
\Big\{ {\cal G} (\xi_1, \lambda_1), \, {\cal G} (\xi_2, \lambda_2) \Big\} = -{\cal G}\Big( \Big[(\xi_1,\lambda_1),(\xi_2,\lambda_2) \Big]_{\cal C} \Big) \, .
\end{equation}
The Courant bracket represents the self T-dual invariant extension of the Lie bracket. 

In the coordinate-free notation, the Courant bracket can be written as
\begin{equation} \label{eq:cfcourant}
\Big[(\xi_1,\lambda_1), (\xi_2,\lambda_2) \Big]_{\cal C} = \Big( [\xi_1,\xi_2]_L, \ {\cal{L}}_{\xi_1} \lambda_2 - {\cal{L}}_{\xi_2} \lambda_1 - \frac{1}{2}d (i_{\xi_1} \lambda_2 - i_{\xi_2} \lambda_1) \Big) \, ,
\end{equation}
with $i_\xi$ being the interior product along the vector field $\xi$, and $d$ being the exterior derivative. The Lie derivative ${\cal L}_\xi$ can be written as their anticommutator 
\begin{equation} \label{eq:Lieder}
{\cal L}_{\xi} =i_{\xi} d + d i_{\xi} \, .
\end{equation}

The Courant bracket does not satisfy the Jacobi identity. Nevertheless, the Jacobiator of the Courant bracket is an exact 1-form \cite{gualtieri}
\begin{equation}
\Big[(\xi_1, \lambda_1), \Big[(\xi_2, \lambda_2), (\xi_3, \lambda_3)\Big]_{\cal C}\Big]_{\cal C} + cycl. = d \varphi, \ \ (d\varphi)_\mu = \partial_\mu \varphi \, .
\end{equation}
However, if one makes the following change of parameters $\lambda_\mu \to \lambda_\mu + \partial_\mu \varphi$, the generator (\ref{eq:gltilde}) does not change 
\begin{equation} \label{eq:reducible}
{\cal G} (\xi, \lambda + \partial \varphi) = {\cal G} (\xi, \lambda) + \kappa \int_0^{2\pi}\varphi^\prime d\sigma = {\cal G} (\xi, \lambda) \, ,
\end{equation} 
since the total derivative integral vanishes for the closed string. Therefore, the deviation from Jacobi identity contributes to the trivial symmetry, and we say that the symmetry is reducible. 

The theory with the metric tensor was already discussed in \cite{dft}, where it was proven that the invariance under both diffeomorphisms and dual diffeomorphisms requires the introduction of the Kalb-Ramond field. In our approach, if we want to include  the T-dual of the general coordinate transformation in the same theory, we obtain the local gauge transformation that constitutes a trivial symmetry, since $\delta_\lambda G_{\mu \nu} = 0$ \cite{doucou,dualsim}. Therefore, it is necessary to include the Kalb-Ramond field, in order to have non-trivial local gauge transformations, which we do in the next section. 

\section{Bosonic string moving in the background characterized by the metric field and the Kalb-Ramond field}
\cleq

In this chapter, we extend the Hamiltonian so that it includes the antisymmetric Kalb-Ramond field. It is possible to obtain this Hamiltonian from the transformation of the generalized metric $G_{MN}$ (\ref{eq:GMN}) under the so called B-transformations. The B-transformations (or B-shifts) \cite{gualtieri} are realized by $e^{\hat{B}}$ , where
\begin{equation} \label{eq:bhat}
\hat{B}^M_{\ N} = 
\begin{pmatrix}
0 & 0 \\
2B_{\mu \nu} & 0 \\ 
\end{pmatrix}.
\end{equation}
As a result of $\hat{B}^2 = 0$, the full transformation is easily obtained 
\begin{equation} \label{eq:ebhat}
(e^{\hat{B}})^M_{\ N} = \begin{pmatrix}
\delta^\mu_\nu & 0 \\
2B_{\mu \nu} & \delta^\nu_\mu
\end{pmatrix} \, .
\end{equation}
Its transpose is given by
\begin{equation} \label{eq:ebminus}
( (e^{\hat{B}})^T)_M^{\ N} = \begin{pmatrix}
\delta^\nu_\mu& -2B_{\mu \nu} \\
0 & \delta^\mu_\nu 
\end{pmatrix} \, ,
\end{equation}
from which it is easy to verify that
\begin{equation} \label{eq:eOe}
( (e^{\hat{B}})^T)_M^{\ K}\ \eta_{KL}\ (e^{\hat{B}})^L_{\ N} = \eta_{MN} \, , 
\end{equation}
meaning they are the elements of the $O(D,D)$ group.

The transformation of generalized metric $G_{MN}$ (\ref{eq:GMN}) under the B-shifts is given by
\begin{equation} \label{eq:HMNC}
G_{MN}  \to( (e^{\hat{B}})^T)_M^{\ K}\ G_{KQ}\ (e^{\hat{B}})^Q_{\ N} \equiv H_{MN} \, ,
\end{equation}
where $H_{MN}$ is the generalized metric
\begin{equation} \label{eq:genmet}
H_{MN} = 
\begin{pmatrix}
G^E_{\mu \nu} & - 2B_{\mu \rho} (G^{-1})^{\rho \nu} \\
2(G^{-1})^{\mu \rho} B_{\rho \nu} & (G^{-1})^{\mu \nu}
\end{pmatrix} \, ,
\end{equation}
and $G^E_{\mu \nu}$ is the effective metric perceived by the open strings, given by
\begin{equation} \label{eq:Geff}
G^E_{\mu \nu} = G_{\mu \nu} - 4 (B G^{-1} B)_{\mu \nu} \, .
\end{equation}
It is straightforward to write the canonical Hamiltonian
\begin{eqnarray} \label{eq:Hcdef}
{\cal  \hat{H}}_C &=& \frac{1}{2\kappa} (X^T)^M H_{MN} X^N \\ \notag
&=& \frac{1}{2\kappa} \pi_\mu (G^{-1})^{\mu \nu}\pi_\nu + \frac{\kappa}{2} x^{\prime \mu} G^E_{\mu \nu} x^{\prime \nu}- 2  x^{\prime \mu} B_{\mu \rho} (G^{-1})^{\rho \nu}  \pi_\nu  \, ,
\end{eqnarray}
as well as the Lagrangian in the canonical form
\begin{eqnarray} \label{eq:langr1}
{\cal  \hat{L}} (\dot{x}, x^\prime, \pi) &=&  \pi_\mu \dot{x}^\mu - {\cal  \hat{H}}_C ( x^\prime, \pi)  \\ \notag
&=& \pi_\mu \dot{x}^\mu
-\frac{1}{2\kappa} \pi_\mu (G^{-1})^{\mu \nu}\pi_\nu - \frac{\kappa}{2} x^{\prime \mu} G^E_{\mu \nu} x^{\prime \nu} + 2  x^{\prime \mu} B_{\mu \rho} (G^{-1})^{\rho \nu}  \pi_\nu .
\end{eqnarray}
On the equations of motion for $\pi_\mu$, we obtain
\begin{equation} \label{eq:pihatdef}
\pi_\mu = \kappa G_{\mu \nu} \dot{x}^\nu - 2\kappa B_{\mu \nu} x^{\prime \nu} \, .
\end{equation}
Substituting (\ref{eq:pihatdef}) into (\ref{eq:langr1}) we find the well known expression for bosonic string Lagrangian \cite{action,regal}
\begin{eqnarray} \label{eq:langr}
{\cal  \hat{L}} (\dot{x}, x^\prime) &=&  \frac{\kappa}{2} \dot{x}^\mu
G_{\mu \nu} \dot{x}^\nu - \frac{\kappa}{2} x^{\prime \mu} G_{\mu \nu} x^{\prime \nu} - 2 \kappa \dot{x}^\mu B_{\mu \nu} x^{\prime \nu} = \kappa \partial_+ x^\mu \Pi_{+ \mu \nu} \partial_-x^\nu \, ,\\ \notag
&& \Pi_{\pm \mu \nu} = B_{\mu \nu} \pm \frac{1}{2} G_{\mu \nu} \, , \ \partial_\pm x^\mu = \dot{x}^\mu \pm x^{\prime \mu} \, .
\end{eqnarray}

It is possible to rewrite the canonical Hamiltonian (\ref{eq:Hcdef}) in terms of the generalized metric $G_{MN}$, that characterizes background with the metric only tensor. Substituting (\ref{eq:HMNC}) into (\ref{eq:Hcdef}), we obtain
\begin{equation}
{\cal  \hat{H}}_C = \frac{1}{2\kappa}(X^T)^M ( (e^{\hat{B}})^T)_M^{\ K} \ G_{KL}\ (e^{\hat{B}})^L_{\ N} X^N =\frac{1}{2 \kappa}  (\hat{X}^T)^M\ G_{MN}\ \hat{X}^N \, , 
\end{equation}
where 
\begin{equation} \label{eq:XXB}
\hat{X}^M = (e^{\hat{B}})^M_{\ N}\ X^N = 
\begin{pmatrix}
\kappa x^{\prime \mu} \\
\pi_\mu + 2 \kappa B_{\mu \nu} x^{\prime \nu}  
\end{pmatrix} \equiv
\begin{pmatrix}
\kappa x^{\prime \mu} \\
i_\mu
\end{pmatrix}
\, ,
\end{equation}
with $i_\mu$ being the auxiliary current, given by
\begin{eqnarray}\label{eq:idef}
i_\mu = \pi_\mu + 2 \kappa B_{\mu\nu} x^{\prime\nu} \,  .
\end{eqnarray}
The algebra of auxiliary currents $i_\mu$ gives rise to the $H$-flux \cite{crdual}
\begin{equation} \label{eq:iialgebra}
\{ i_\mu (\sigma), i_\nu (\bar{\sigma}) \} = - 2\kappa B_{\mu \nu \rho} x^{\prime \rho} \delta (\sigma - \bar{\sigma}) \, ,
\end{equation}
where the structural constants are the Kalb-Ramond field strength components, given by 
\begin{equation} \label{eq:bmnr}
B_{\mu \nu \rho} = \partial_\mu B_{\nu \rho} + \partial_\nu B_{\rho \mu} + \partial_\rho B_{\mu \nu} \, .
\end{equation} 

\subsection{Symmetry generator}

Let us extend the symmetry transformations of the background fields for the theory with the non-trivial Kalb-Ramond field. The infinitesimal general coordinate transformations of the background fields are given by 
\begin{equation} \label{eq:transf1}
\delta_\xi G_{\mu \nu} = {\cal L}_\xi G_{\mu \nu} \, ,\ \delta_\xi B_{\mu \nu} = {\cal L}_\xi B_{\mu \nu} \, ,
\end{equation}
where the action of the Lie derivative ${\cal L}_\xi$ (\ref{eq:Lieder}) on the Kalb-Ramond field is given by \cite{dualsim}
\begin{equation}
{\cal L}_\xi B_{\mu \nu} = \xi^\rho \partial_\rho B_{\mu \nu} + \partial_\mu \xi^\rho B_{\rho \nu} - \partial_\nu \xi^\rho B_{\rho \mu} \, ,
\end{equation}
while its action on the metric field is the same as in (\ref{eq:LieG}). The local gauge transformations of the background fields are \cite{dualsim}
\begin{equation} \label{eq:transf2}
\delta_\lambda G_{\mu \nu} = 0 \, , \ \delta_\lambda B_{\mu \nu} = \partial_\mu \lambda_\nu - \partial_\nu \lambda_\mu \, .
\end{equation}

Rewriting the symmetry generator ${\cal G}(\xi,\lambda)$ (\ref{eq:gltilde}) in terms of the basis defined by components of $\hat{X}^M$ (\ref{eq:XXB}), one obtains
\begin{eqnarray} \label{eq:GC}\notag
{\cal G}(\xi,\lambda)&& = \int d\sigma (\Lambda^T)^M \eta_{MN} X^N = \int d\sigma (\hat{\Lambda}^T)^M ((e^{-\hat{B}})^T)^{\ K}_{M} \eta_{KL} (e^{-\hat{B}})^L_{\ N} \hat{X}^N \\ 
&&= \int d\sigma (\hat{\Lambda}^T)^M \eta_{MN} \hat{X}^N \, ,
\end{eqnarray}
where (\ref{eq:eOe}) was used in the last step, and $\hat{\Lambda}^M$ is a new double gauge parameter, given by
\begin{equation}\label{eq:LBdouble}
\hat{\Lambda}^M = (e^{\hat{B}})^M_{\ N} \Lambda^N = \begin{pmatrix}
\delta^\mu_\nu & 0 \\
2B_{\mu \nu} & \delta^\nu_\mu
\end{pmatrix}
\begin{pmatrix}
\xi^\nu \\
\lambda_\nu
\end{pmatrix} = 
\begin{pmatrix}
\xi^\mu \\
\lambda_\mu + 2 B_{\mu \nu} \xi^\nu
\end{pmatrix}
\equiv
\begin{pmatrix}
\xi^\mu \\
\hat{\lambda}_\mu 
\end{pmatrix}
\, .
\end{equation}
We are going to mark the right hand side of (\ref{eq:GC}) as a new generator 
\begin{equation} \label{eq:GBdef}
{\cal G}^{\hat{B}} (\xi, \hat{\lambda}) = \int d\sigma \Big[ \xi^\mu i_\mu + \hat{\lambda}_\mu \kappa x^{\prime \mu} \Big] \, ,
\end{equation}
which equals the generator (\ref{eq:gltilde}), when the relations between the gauge parameters (\ref{eq:LBdouble}) are satisfied ${\cal G}(\xi,\hat{\lambda} - 2B_{\mu \nu} \xi^\nu ) = {\cal G}^{\hat{B}} (\xi, \hat{\lambda})$. The expression (\ref{eq:GBdef}) exactly corresponds to the symmetry generator obtained in \cite{dualsim}, where $\xi^\mu$ are parameters of general coordinate transformations and $\hat{\lambda}_\mu$ are parameters of local gauge transformations, that respectively correspond to transformations of the background fields  (\ref{eq:transf1}) and (\ref{eq:transf2}).

Our goal is to obtain the algebra in the form
\begin{equation} \label{eq:GBprovalg}
\Big\{ {\cal G}^{\hat{B}} (\xi_1, \hat{\lambda}_1), \, {\cal G}^{\hat{B}} (\xi_2, \hat{\lambda}_2) \Big\} = -{\cal G}^{\hat{B}}(\xi,\hat{\lambda})\, ,
\end{equation}
where 
\begin{equation} \label{eq:llhat}
\lambda_{i \mu} = \hat{\lambda}_{i \mu }- 2B_{\mu \nu} \xi_i^\nu \, ,\ i = 1,2\, ;\ \ \lambda_\mu = \hat{\lambda}_\mu - 2B_{\mu \nu} \xi^\nu \, ,
\end{equation}
due to (\ref{eq:LBdouble}). The Poisson bracket between canonical variables (\ref{eq:PBR}) remains the same after the introduction of the Kalb-Ramond field. Therefore the results from previous chapter, as well as mutual relations between coefficients in different bases can be used to obtain the algebra (\ref{eq:GBprovalg}). Firstly, substituting (\ref{eq:llhat}) into the second equation in (\ref{eq:courantt}), one obtains
\begin{eqnarray} \label{eq:LLB1}
\lambda_\mu &=&\xi_1^\nu (\partial_\nu \hat{\lambda}_{2 \mu} - \partial_\mu \hat{\lambda}_{2 \nu}) - \xi_2^\nu (\partial_\nu \hat{\lambda}_{1 \mu} - \partial_\mu \hat{\lambda}_{1 \nu}) +\frac{1}{2} \partial_\mu (\xi_1 \hat{\lambda}_2- \xi_2 \hat{\lambda}_1 )\\ \notag
&&+ 2 B_{\mu \nu \rho} \xi^\nu_1 \xi^\rho_2 -2 B_ {\mu \nu} (\xi^\rho_1 \partial_\rho \xi_2^\nu - \xi^\rho_2 \partial_\rho \xi_1^\nu) \, .
\end{eqnarray}
Secondly, substituting the previous equation in (\ref{eq:LBdouble}), one obtains
\begin{eqnarray} \label{eq:xiLB}
\xi^\mu &=& \xi_1^\nu \partial_\nu \xi_2^\mu - \xi_2^\nu \partial_\nu \xi_1^\mu \, ,\\ \notag 
\hat{\lambda}_\mu &=&\xi_1^\nu (\partial_\nu \hat{\lambda}_{2 \mu} - \partial_\mu \hat{\lambda}_{2 \nu}) - \xi_2^\nu (\partial_\nu \hat{\lambda}_{1 \mu} - \partial_\mu \hat{\lambda}_{1 \nu}) +\frac{1}{2} \partial_\mu (\xi_1 \hat{\lambda}_2- \xi_2 \hat{\lambda}_1 )+ 2  B_{\mu \nu \rho} \xi^\nu_1 \xi^\rho_2 \, .
\end{eqnarray} 
The above relations define the twisted Courant bracket $[(\xi_1,\hat{\lambda}_1),(\xi_2,\hat{\lambda}_2)]_{{\cal C}_B} = (\xi,\hat{\lambda})$ \cite{twist}. This is the bracket of the symmetry transformations 
\begin{equation}
\Big\{ {\cal G}^{\hat{B}} (\xi_1, \hat{\lambda}_1), \, {\cal G}^{\hat{B}} (\xi_2, \hat{\lambda}_2) \Big\} = -{\cal G}^{\hat{B}}\Big( \Big[(\xi_1,\hat{\lambda}_1),(\xi_2,\hat{\lambda}_2)\Big]_{ {\cal C}_B}  \Big)\, ,
\end{equation}
in the theory defined by both metric and Kalb-Ramond field. 

In the coordinate free notation, the twisted Courant bracket is given by
\begin{equation} \label{eq:CBbracket}
\Big[ (\xi_1,\hat{\lambda}_1), (\xi_2,\hat{\lambda}_2) \Big]_{{\cal C}_B}  = \Big( [\xi_1,\xi_2]_L, {\cal{L}}_{\xi_1} \hat{\lambda}_2 - {\cal{L}}_{\xi_2} \hat{\lambda}_1 - \frac{1}{2}d (i_{\xi_1} \hat{\lambda}_2 - i_{\xi_2} \hat{\lambda}_1) + H(\xi_1,\xi_2,.) \Big) \, ,
\end{equation}
where $H(\xi_1,\xi_2,.)$ represents the contraction of the $H$-flux $H = d B$ (\ref{eq:bmnr}) with two gauge parameters $\xi_1$ and $\xi_2$. This term is the corollary of the non-commutativity of the auxiliary currents $i_\mu$ (\ref{eq:iialgebra}), due to twisting of the Courant bracket with the Kalb-Ramond field. In special case when the Kalb-Ramond field $B$ is a closed form $d B= 0$, the twisted Courant bracket (\ref{eq:CBbracket}) reduces to the Courant bracket (\ref{eq:cfcourant}). This can also be seen from the well known fact that $B$-shifts (\ref{eq:ebhat}) are symmetries of the Courant bracket when $B$ is a closed form \cite{gualtieri}.

\section{Courant bracket twisted by $\theta^{\mu \nu}$}
\cleq

When both the metric and the Kalb-Ramond field are present in the theory, the expressions for T-dual fields are given by \cite{buscher}
\begin{equation}\label{eq:tdbf}
^\star G^{\mu\nu} =
(G_{E}^{-1})^{\mu\nu}, \quad
^\star B^{\mu\nu} =
\frac{\kappa}{2}
{\theta}^{\mu\nu}  \, ,
\end{equation}
where ${\theta}^{\mu\nu}$ is the non-commutativity parameter for the string endpoints on a D-brane \cite{witten}, given by
\begin{equation} \label{eq:thetadef}
\theta^{\mu \nu} = -\frac{2}{\kappa} (G^{-1}_E)^{\mu \rho}  B_{\rho \sigma} (G^{-1})^{\sigma \nu} \, .
\end{equation}
We say that two quantities are self T-dual, if they are invariant under the interchange \cite{crdual}
\begin{equation} \label{eq:selfTdual}
\pi_\mu \leftrightarrow \kappa x^{\prime \mu} \, , \ G_{\mu \nu} \leftrightarrow (G_{E}^{-1})^{\mu\nu} \, , \ B_{\mu \nu} \leftrightarrow \frac{\kappa}{2}\theta^{\mu \nu} \, .
\end{equation}
When the Kalb-Ramond field is set to zero $B_{\mu \nu} = 0$, (\ref{eq:selfTdual})  reduces to the self T-duality transformation laws in the background without the $B$ field (\ref{eq:xpidual}). 

From the relations (\ref{eq:selfTdual}), it is apparent that the introduction of Kalb-Ramond field breaks down the self T-duality invariance of the symmetry generator (\ref{eq:GBdef}). To find a new self T-dual invariant generator, we will analogously to the prior construction start with the background containing only T-dual metric. The Hamiltonian in the metric only background, similar to (\ref{eq:HG}), reads
\begin{equation}
^\star{\cal H}_{C} = \frac{1}{2 \kappa} \pi_\mu (G_E^{-1})^{\mu \nu} \pi_\nu + \frac{\kappa}{2} x^{\prime \mu} G^E_{\mu \nu} x^{\prime \nu} = (X^T)^M \ {^\star G}_{MN} X^N \, ,
\end{equation}
where $^\star G_{MN}$ is the T-dual generalized metric for the above Hamiltonian, given by
\begin{equation} \label{eq:Gstar}
^\star G_{MN} = 
\begin{pmatrix}
G^E_{\mu \nu} & 0 \\
0 & (G_E^{-1})^{\mu \nu} 
\end{pmatrix} \, .
\end{equation}
Note that the self T-duality is realized as the joint action of the permutation of the coordinate $\sigma$-derivatives with the canonical momenta and the swapping all the fields in (\ref{eq:GMN}) for their T-duals. This is equivalent to the Buscher's procedure \cite{buscher,tdual,tdual1}, when it is done in the same phase space.

In order to construct the Hamiltonian in the self T-dual description, we consider how the T-dual generalized metric (\ref{eq:Gstar}) is transformed with respect to the so called $\theta$-transformations $e^{\hat{\theta}}$, where
\begin{equation}
\hat{ \theta}^M_{\ N} = 
\begin{pmatrix}
0 & \kappa \theta^{\mu \nu} \\
0 & 0 
\end{pmatrix} = 
\begin{pmatrix}
0 & 2 \ {^\star B}^{\mu \nu} \\
0 & 0
\end{pmatrix} \, .
\end{equation}
The full exponential $e^{\hat{\theta}}$ is given by
\begin{equation} \label{eq:enateta}
(e^{\hat{\theta}})^M_{\ N} = 
\begin{pmatrix}
 \delta^\mu_\nu & \kappa \theta^{\mu \nu} \\
0 & \delta^\nu_\mu
\end{pmatrix} \, ,
\end{equation}
and its transpose by
\begin{equation} \label{eq:eminteta}
 ( (e^{\hat{\theta}})^T )_{M}^{\ N} =
\begin{pmatrix}
\delta^\nu_\mu & 0\\
-\kappa \theta^{\mu \nu}  & \delta^\mu_\nu  
\end{pmatrix} \, .
\end{equation}
They are elements of the $O(D,D)$ group as well, i.e.
\begin{equation} \label{eq:etOet}
( (e^{\hat{\theta}})^T )_{M}^{\ L}\ \eta_{LK}\ (e^{\hat{\theta}})^K_{\ N} = \eta_{MN} \, .
\end{equation}
Under (\ref{eq:enateta}), the T-dual generalized metric (\ref{eq:Gstar}) transforms in the following way 
\begin{equation} \label{eq:GtoH}
^\star G_{MN} \to \  ( (e^{\hat{\theta}})^T )_{M}^{\ L}\ {^\star G}_{LK} (e^{\hat{\theta}})^K_{\ N} \equiv \ ^\star H_{MN}  \, , 
\end{equation}
where
\begin{equation} \label{eq:GstH}
^\star H_{MN}=
\begin{pmatrix}
G^E_{\mu \nu}& - 2B_{\mu \rho} (G^{-1})^{\rho \nu} \\
 2(G^{-1})^{\mu \rho} B_{\rho \nu} & (G^{-1})^{\mu \nu}
\end{pmatrix} \, ,
\end{equation}
which is exactly equal to the generalized metric (\ref{eq:genmet}). From it we can write the T-dual Hamiltonian 
\begin{eqnarray} \label{eq:dualH}
^\star {\cal H}_C  &=& \frac{1}{2\kappa} (X^T)^M\ ^\star H_{MN} X^N \\ \notag
&=& \frac{1}{2\kappa} \pi_\mu (G^{-1})^{\mu \nu}\pi_\nu + \frac{\kappa}{2} x^{\prime \mu} G^E_{\mu \nu} x^{\prime \nu}- 2  x^{\prime \mu} B_{\mu \rho} (G^{-1})^{\rho \nu}  \pi_\nu \equiv {\cal \hat{H}}_C  \, .
\end{eqnarray}
The canonical Lagrangian is given by
\begin{eqnarray} \label{eq:dlagr1}
^\star {\cal L}(\pi,\dot{x},x) &=&  \pi_\mu \dot{x}^\mu - {^\star {\cal  H}}_C ( x^\prime, \pi)  \\ \notag
&=& \pi_\mu \dot{x}^\mu
-\frac{1}{2\kappa} \pi_\mu (G^{-1})^{\mu \nu}\pi_\nu - \frac{\kappa}{2} x^{\prime \mu} G^E_{\mu \nu} x^{\prime \nu} + 2  x^{\prime \mu} B_{\mu \rho} (G^{-1})^{\rho \nu}  \pi_\nu \, ,
\end{eqnarray}
from which one easily obtains 
\begin{equation} \label{eq:pidual}
\pi_\mu = \kappa G_{\mu \nu} \dot{x}^\nu - 2\kappa B_{\mu \nu} x^{\prime \nu} \, .
\end{equation}
We see that the canonical momentum remains the same, which is expected, since the self T-duality is realized in the same phase space.
Substituting (\ref{eq:pidual}) into (\ref{eq:dlagr1}), one obtains
\begin{equation} \label{eq:dlagr}
^\star {\cal L}(\dot{x},x) = \frac{\kappa}{2} \dot{x}^\mu
G_{\mu \nu} \dot{x}^\nu - \frac{\kappa}{2} x^{\prime \mu} G_{\mu \nu} x^{\prime \nu} - 2 \kappa \dot{x}^\mu B_{\mu \nu} x^{\prime \nu} = \kappa \partial_+ x^\mu \Pi_{+ \mu \nu} \partial_-x^\nu  \, .
\end{equation}
It is obvious that both the Hamiltonian and the Lagrangian are invariant under the self T-duality.

In the same manner as in the previous chapter, substituting (\ref{eq:GtoH}) into (\ref{eq:dualH}), we rewrite the Hamiltonian
\begin{equation}
^\star{\cal \hat{H}}_C = \frac{1}{2 \kappa} (X^T)_M^{\ L}\ ((e^{\hat{\theta}})^T)^{\ K}_{L}\ {^\star G}_{KJ}\ (e^{\hat{\theta}})^J_{\ N}\ X^N =\frac{1}{2 \kappa}  \tilde{X}^M {^\star G}_{MN} \tilde{X}^N \, ,
\end{equation}
where 
\begin{equation}
\tilde{X}^M = (e^{\hat{\theta}})^M_{\ N} X^N = 
\begin{pmatrix}
 \delta^\mu_\nu & \kappa \theta^{\mu \nu} \\
0 & \delta^\nu_\mu
\end{pmatrix}
\begin{pmatrix}
\kappa x^{\prime \nu} \\
\pi_\nu
\end{pmatrix} = 
\begin{pmatrix}
\kappa x^{\prime \mu} + \kappa \theta^{\mu \nu} \pi_\nu \\
\pi_\mu
\end{pmatrix} \equiv
\begin{pmatrix}
k^\mu \\
\pi_\mu 
\end{pmatrix} \, ,
\end{equation}
and $k^\mu$ is an auxiliary current, given by
\begin{eqnarray}\label{eq:defk}
k^\mu = \kappa  x^{\prime\mu}    + \kappa  \theta^{\mu\nu} \pi_\nu   \,  .
\end{eqnarray}
The Poisson bracket algebra of these currents is obtained in \cite{crdual}
\begin{equation} \label{eq:pbrk}
\{ k^\mu (\sigma), k^{\nu} (\bar{\sigma}) \} = -\kappa Q_\rho^{\ \mu \nu} k^\rho \delta(\sigma - \bar{\sigma}) - \kappa^2 R^{\mu \nu \rho} \pi_\rho \delta (\sigma - \bar{\sigma}) \, ,
\end{equation}
where $Q$ and $R$ are non-geometric fluxes \cite{stw}, given by
\begin{equation} \label{eq:QRdef}
Q_{\rho}^{\ \mu \nu} = \partial_\rho \theta^{\mu \nu}, \ \ \ \ \ R^{\mu \nu \rho} = \theta^{\mu \sigma} \partial_\sigma \theta^{\nu \rho} + \theta^{\nu \sigma} \partial_\sigma \theta^{\rho \mu} + \theta^{\rho \sigma} \partial_\sigma \theta^{\mu \nu} \, .
\end{equation}

We now define a new double gauge parameter 
\begin{equation} \label{eq:tildeL}
\tilde{\Lambda}^M = (e^{\hat{\theta}})^M_{\ N} \Lambda^N = 
\begin{pmatrix}
\delta^\mu_\nu & \kappa \theta^{\mu \nu} \\
0 & \delta^\nu_\mu
\end{pmatrix}
\begin{pmatrix}
\xi^\nu \\
\lambda_\nu
\end{pmatrix} = 
\begin{pmatrix}
\xi^\mu + \kappa \theta^{\mu \nu} \lambda_\nu \\
\lambda_\mu 
\end{pmatrix} \equiv
\begin{pmatrix}
\hat{\xi}^\mu \\
\lambda_\mu 
\end{pmatrix} \, .
\end{equation}
The generator (\ref{eq:gltilde}) written in terms of new gauge parameters ${\cal G}(\hat{\xi}-\kappa \theta \lambda, \lambda)\equiv{\cal G^{\hat{\theta}}} (\hat{\xi}, \lambda)$ is given by
\begin{equation} \label{eq:Gthetadef}
{\cal G^{\hat{\theta}}} (\hat{\xi} , \lambda)=\int d\sigma\Big[\hat{\xi}^\mu \pi_\mu+ \lambda_\mu k^\mu \Big] \, .
\end{equation}
The auxiliary currents $i_\mu$ (\ref{eq:idef}) and $k^\mu$ (\ref{eq:defk}) are related by the self T-duality relations (\ref{eq:selfTdual}). Moreover, one easily demonstrates that the self T-dual image of the generator ${\cal G}^{\hat{B}}$ (\ref{eq:GBdef}) is the generator ${\cal G}^{\hat{\theta}}$ (\ref{eq:Gthetadef}).

Like in a previous case, we want to obtain the algebra in the form
\begin{equation}
\Big\{ {\cal G^{\hat{\theta}}} (\hat{\xi}_1, \lambda_1), \, {\cal G^{\hat{\theta}}} (\hat{\xi}_2, \lambda_2) \Big\} = -{\cal G^{\hat{\theta}}}(\hat{\xi},\lambda) \, ,
\end{equation}
where from (\ref{eq:tildeL}) we read the relations between the old and new gauge parameters
\begin{equation} \label{eq:xihatxi}
\xi_i^\mu = \hat{\xi}_i^\mu - \kappa \theta^{\mu \nu} \lambda_{i\nu} \, ,\ i=1,2; \ \ \xi^\mu = \hat{\xi}^\mu - \kappa \theta^{\mu \nu} \lambda_\nu \, .
\end{equation}
Combining (\ref{eq:xihatxi}), (\ref{eq:courantt}) and (\ref{eq:tildeL}), one obtains
\begin{eqnarray} \label{eq:xiLR}
\hat{\xi}^\mu &=&\ \hat{\xi}_1^\nu \partial_\nu \hat{\xi}_2^\mu - \hat{\xi}_2^\nu \partial_\nu\hat{\xi}_1^\mu + \\ \notag
&& -\kappa \theta^{\mu \nu}\Big( \hat{\xi}_1^\rho (\partial_\nu \lambda_{2 \rho}-\partial_\rho \lambda_{2 \nu}) - \hat{\xi}_2^\rho ( \partial_\nu \lambda_{1 \rho}-\partial_\rho \lambda_{1 \nu}) -\frac{1}{2} \partial_\nu (\hat{\xi}_1 \lambda_{2} - \hat{\xi}_2 \lambda_1) \Big) \\ \notag
&& + \kappa \hat{\xi}_1^\nu \partial_\nu (\lambda_{2 \rho} \theta^{\rho \mu})-\kappa \hat{\xi}_2^\nu \partial_\nu (\lambda_{1 \rho} \theta^{\rho \mu})+\kappa (\lambda_{1 \nu} \theta^{\nu \rho}) \partial_\rho \hat{\xi}_2^\mu -\kappa (\lambda_{2 \nu}\theta^{\nu \rho}) \partial_\rho \hat{\xi}_1^\mu \\ \notag
&&+\kappa^2 R^{\mu \nu \rho} \lambda_{1 \nu}\lambda_{2 \rho} \, , \\ \notag 
\lambda_\mu &= &\ \hat{\xi}_1^\nu (\partial_\nu \lambda_{2 \mu} - \partial_\mu \lambda_{2 \nu}) - \hat{\xi}_2^\nu (\partial_\nu \lambda_{1 \mu} - \partial_\mu \lambda_{1 \nu}) +\frac{1}{2}\partial_\mu(\hat{\xi}_1 \lambda_2 - \hat{\xi}_2 \lambda_1) \\ \notag
&& + \kappa \theta^{\nu \rho} (\lambda_{1 \nu}\partial_\rho \lambda_{2 \mu}-\lambda_{2 \nu} \partial_\rho \lambda_{1 \mu})+ \kappa \lambda_{1 \rho} \lambda_{2 \nu} Q_\mu^{\ \rho \nu} \, .
\end{eqnarray}
The relations (\ref{eq:xiLR}) define a bracket $[(\hat{\xi}_1,\lambda_1),(\hat{\xi}_2,\lambda_2)]_{ {\cal C}_\theta} = (\hat{\xi},\lambda)$ that is known as the $\theta$-twisted Courant bracket, or Roytenberg bracket. It is related by self T-duality with the twisted Courant bracket, when the relations between the fields (\ref{eq:tdbf}) hold \cite{crdual}.

In the coordinate free notation, the $\theta$-twisted Courant bracket is given by 
\begin{align} \label{eq:cfroytenberg}
[(\hat{\xi}_1,\lambda_1),(\hat{\xi}_2,\lambda_2)]_{ {\cal C}_\theta} = & \Big( [\hat{\xi}_1,\hat{\xi}_2]_L - \kappa [\hat{\xi}_2, \lambda_1 \theta]_L + \kappa [\hat{\xi}_1,\lambda_2 \theta]_L + \frac{\kappa^2}{2} [\theta,\theta]_S(\lambda_1 , \lambda_2, .) \\ \notag
& +\kappa \Big( {\cal{L}}_{\hat{\xi}_2} \lambda_1 -{\cal{L}}_{\hat{\xi}_1} \lambda_2 + \frac{1}{2} d (i_{\hat{\xi}_1} \lambda_2 - i_{\hat{\xi}_2} \lambda_1 ) \Big) \theta, \\ \notag
& {\cal{L}}_{\hat{\xi}_1}\lambda_2 - {\cal{L}}_{\hat{\xi}_2} \lambda_1 -\frac{1}{2} d (i_{\hat{\xi}_1} \lambda_2 - i_{\hat{\xi}_2} \lambda_1 ) - \kappa [\lambda_1 ,\lambda_2]_{ \theta}\Big) \, , \notag
\end{align}
where $ [\theta,\theta]_S(\lambda_1, \lambda_2, .)$ represents the Schouten-Nijenhuis bracket \cite{SNB} contracted with two 1-forms, that when having bi-vectors as domain is given by
\begin{equation} \label{eq:SNb}
\left. [\theta, \theta]_S \right| ^{\mu \nu \rho} = \epsilon^{\mu \nu \rho}_{\alpha \beta \gamma} \theta^{\sigma \alpha} \partial_\sigma \theta^{\beta \gamma} = 3 R^{\mu \nu \rho} \, ,
\end{equation}
where 
\begin{equation}
\epsilon^{\mu \nu \rho}_{\alpha \beta \gamma} = 
\begin{vmatrix}
\delta^\mu_\alpha & \delta^\nu_\beta & \delta^\rho_\gamma \\ 
\delta^\nu_\alpha & \delta^\rho_\beta & \delta^\mu_\gamma \\
\delta^\rho_\alpha & \delta^\mu_\beta & \delta^\nu_\gamma
\end{vmatrix}\, ,
\end{equation}
and $[\lambda_1, \lambda_2]_{\theta}$ is the Koszul bracket \cite{koszul} given by
\begin{equation} \label{eq:koszul}
[\lambda_1, \lambda_2]_\theta = {\cal{L}}_{\theta \lambda_1} \lambda_2 - {\cal{L}}_{\theta \lambda_2} \lambda_1 + d(\theta(\lambda_1, \lambda_2)) \, .
\end{equation}
The Koszul bracket is a generalization of the Lie bracket on the space of differential forms, while the Schouten-Nijenhuis bracket is a generalization of the Lie bracket on the space of multi-vectors.

\section{C-bracket}
\cleq
In this section, we will show how our results can be generalized, so that they give rise to the $C$-bracket \cite{siegel1,siegel} as the T-dual invariant bracket, in the accordance with \cite{doucou}. Consider that T-dual theory is defined in the T-dual phase space, characterized by T-dual coordinates $y_\mu$ and the T-dual momenta ${}^\star \pi^\mu$. They are related with the initial phase space by T-duality relations \cite{buscher}
\begin{eqnarray}\label{eq:TD}
\pi_\mu \simeq \kappa y^{\prime}_\mu \, , \qquad {}^\star \pi^\mu \simeq \kappa x^{\prime \mu} \, .
\end{eqnarray}
We can define a double phase space obtained as a sum of two canonical phase spaces. Let us introduce the double coordinate
\begin{equation}
X^M = \begin{pmatrix}
x^\mu \\
y_\mu \\
\end{pmatrix} \, ,
\end{equation}
as well as the double canonical momentum
\begin{equation} \label{eq:PiM}
\Pi_M = \begin{pmatrix}
\pi_\mu \\
{}^\star \pi^\mu
\end{pmatrix} \, .
\end{equation}
In this notation, the T-duality laws (\ref{eq:TD}) take a form
\begin{eqnarray}\label{eq:TD1}
\Pi_M  \simeq \kappa \, \eta_{M N}  X^{\prime M}  \, ,
\end{eqnarray}
where $\eta_{MN}$ is the $O(D,D)$ metric (\ref{eq:Omega}).

\subsection{Poisson brackets of canonical variables}

The standard Poisson bracket algebra is assumed for both initial and T-dual phase space
\begin{eqnarray}\label{eq:}
\{x^\mu (\sigma), \pi_\nu ({\bar \sigma}) \} = \delta^\mu_\nu \delta (\sigma - {\bar \sigma} ) \, , \qquad
\{y_\mu (\sigma), {}^\star \pi^\nu ({\bar \sigma}) \} = \delta_\mu^\nu \delta (\sigma - {\bar \sigma} ) \, ,
\end{eqnarray}
with other bracket of canonical variables within the same phase space being zero. 

 For the remaining Poisson bracket relations, one must use the consistency with T-duality relations. Firstly, applying the T-dualization along all initial coordinates $x^\mu$, i.e. the second relation of (\ref{eq:TD}) on the Poisson bracket algebra between coordinates derivatives, one obtains
\begin{equation}
\{ \kappa x^{\prime \mu} (\sigma),  \kappa y_\nu^{\prime} (\bar{\sigma}) \} \simeq \{ {}^\star \pi^\mu (\sigma),  \kappa y_\nu^{\prime} (\bar{\sigma}) \} = \kappa \delta^\mu_\nu \delta^{\prime}(\sigma-\bar{\sigma}) \, . 
\end{equation}
Similarly, applying the T-dualization along all T-dual coordinates $y_\mu$, i.e. the first relation of (\ref{eq:TD}), one obtains
\begin{equation}
\{ \kappa x^{\prime \mu} (\sigma),  \kappa y_\nu^{\prime} (\bar{\sigma}) \} \simeq \{ \kappa x^{\prime \mu} (\sigma),  \pi_\nu(\bar{\sigma}) \} = \kappa \delta^\mu_\nu \delta^{\prime}(\sigma-\bar{\sigma}) \, . 
\end{equation}
Hence, we conclude
\begin{equation}
\{ \kappa x^{\prime \mu} (\sigma), \kappa y^\prime_\nu (\bar{\sigma})\} = \kappa \delta^\mu_\nu \delta^{\prime} (\sigma-\bar{\sigma}) \, .
\end{equation}
The successive integration along both $\sigma$ and $\bar{\sigma}$ for the appropriate choice of the integration constant produces the relation \cite{cannc}
\begin{equation}
\{ \kappa x^\mu (\sigma), \kappa y_\nu (\bar{\sigma}) \} = -\kappa \delta^\mu_\nu \theta(\sigma-\bar{\sigma}) \, ,
\end{equation}
where  
\begin{equation}
\theta(\sigma) = 
     \begin{cases}
       -\frac{1}{2} &\quad \sigma = -\pi \\
       0 &\quad -\pi < \sigma < \pi \\
       \frac{1}{2} &\quad \sigma = \pi\\
     \end{cases} \, .
\end{equation}	

Secondly, taking into the account T-duality (\ref{eq:TD}), the Poisson bracket algebra of momenta is easily transformed into 
\begin{eqnarray}\label{eq:1}
\{ \pi_\mu (\sigma) , {}^\star \pi^\nu ({\bar \sigma}) \} \simeq \kappa \{ \pi_\mu (\sigma) , x^{\prime \nu} ({\bar \sigma}) \} =
\kappa \delta_\mu^\nu \delta^\prime (\sigma - {\bar \sigma} ) \, ,
\end{eqnarray}
when T-dualization is applied along the coordinates $y_\mu$, and
\begin{eqnarray}\label{eq:2}
\{ \pi_\mu (\sigma) , {}^\star \pi^\nu ({\bar \sigma}) \} \simeq \kappa \{ y^\prime_\mu (\sigma) , {}^\star \pi^\nu  ({\bar \sigma}) \} =
\kappa \delta_\mu^\nu \delta^\prime (\sigma - {\bar \sigma} ) \, ,
\end{eqnarray}
when it is applied along the coordinates $x^\mu$. As in the previous case, we obtain
\begin{eqnarray}\label{eq:3}
\{ \pi_\mu (\sigma) , {}^\star \pi^\nu ({\bar \sigma}) \}  = \kappa \delta_\mu^\nu \delta^\prime (\sigma - {\bar \sigma} ) \, .
\end{eqnarray}
In a same manner, it is easy to demonstrate that 
\begin{equation}
\{ x^\mu (\sigma), {}^\star \pi^\nu ({\bar \sigma}) \} = 0 \, ,\ \{ y_\mu (\sigma), \pi_\nu ({\bar \sigma}) \} = 0 \, .
\end{equation}

In a double space, the above relations can be simply written as
\begin{equation} \label{eq:PBPX}
\{\kappa X^M (\sigma), \kappa X^N ({\bar \sigma}) \} = -\kappa \eta^{MN} \theta(\sigma - {\bar \sigma} ) \, , \qquad
\{ \Pi_M (\sigma), \Pi_N ({\bar \sigma}) \} = \kappa \, \eta_{M N} \delta^\prime (\sigma - {\bar \sigma} ) \, .
\end{equation}

\subsection{Generator in double space}
Now let us extend the generator of general coordinate transformations, so that it includes the T-dual version of that generator
\begin{equation}\label{eq:Gen1}
G(\xi, \lambda) = \int d\sigma{\cal G} (\xi, \lambda) = \int d\sigma  \Big[ \xi^\mu (x, y)  \pi_\mu  + \lambda_\mu (x, y) {}^\star \pi^\mu \Big] \,  ,
\end{equation}
where the symmetry parameters $\xi$ and $\lambda$ depend on both initial coordinates $x^\mu$ and T-dual coordinates $y_\mu$.
The generator ${\cal G} (\xi, \lambda)$ can be rewritten in terms of double canonical variables as 
\begin{equation}\label{eq:Gen}
 {\cal G} (\Lambda) =  \Lambda^M (x, y) \eta_{M N} \Pi^N  \, \qquad \Longleftrightarrow \qquad {\cal G}_\Lambda  = \langle \Lambda, \Pi   \rangle \,  ,
\end{equation}
where
\begin{equation}
\Lambda^M  (X) = 
\begin{pmatrix}
\xi^\mu  (x^\mu, y_\mu) \\
\lambda_\mu (x^\mu, y_\mu) 
\end{pmatrix} \, .
\end{equation}
This generator is manifestly $O(D, D)$ invariant. 

We are interested in the algebra of the form
\begin{equation} \label{eq:GGCbr}
\{G(\Lambda_1), G(\Lambda_2) \} = - G (\Lambda) \, .
\end{equation}
To obtain it, it is convenient to introduces double derivative 
 \begin{eqnarray}
\partial_M =  \begin{pmatrix}
  \partial_\mu  \\
 \tilde{\partial}^\mu  \\
 \end{pmatrix}  \,   \qquad  \Big( \partial_\mu \equiv \frac{\partial}{\partial x^\mu} \, , \quad  \tilde{\partial}^\mu \equiv \frac{\partial}{\partial y_\mu}  \Big) \, ,
\end{eqnarray}
so that the following Poisson bracket relations can be written
\begin{eqnarray}\label{eq:PBL}
\{\Lambda^M (\sigma), \Pi_N ({\bar \sigma}) \} = \partial_N \Lambda^M  \delta (\sigma - {\bar \sigma} ), 
\{ \Lambda^M (\sigma), \Lambda^N ({\bar \sigma}) \} = -\frac{1}{\kappa} \, \partial^P \Lambda_1^M \partial_P \Lambda_2^N \theta(\sigma - {\bar \sigma} ) \, .
\end{eqnarray}
The second relation makes the situation more complicated, since it would result in the symmetry algebra not closing on another generator. However, in the accordance with \cite{doucou,siegel}, we can consider restricted parameters on isotropic spaces, for which $\Delta = \eta^{PQ} \partial_P \partial_Q = \partial^Q \partial_Q$ anihilates all gauge parameters, as well as their products. Therefore, we write
\begin{equation}
\Delta (\Lambda_1^M \Lambda_2^N) = \Delta \Lambda_1^M \Lambda_2^N + 2\partial_Q \Lambda_1^M \partial^Q \Lambda_2^N + \Lambda_1^M \Delta \Lambda_2^N = 0 \, ,
\end{equation}
from which one obtains
\begin{equation} \label{eq:restr}
\partial_Q \Lambda_1^M \partial^Q \Lambda_2^N = 0 \, .
\end{equation}
Substituting (\ref{eq:restr}) into (\ref{eq:PBL}), we obtain
\begin{equation} \label{eq:LL}
\{ \Lambda^M (\sigma), \Lambda^N ({\bar \sigma}) \} = 0 \, .
\end{equation}
We see that the restriction of gauge parameters to isotropic spaces is necessary for the algebra of generator (\ref{eq:Gen1}) to be closed. 

Now we are ready to calculate the algebra. Using the second relation of (\ref{eq:PBPX}), the first relation of (\ref{eq:PBL}), and (\ref{eq:LL}), we have
\begin{equation}\label{eq:}
\{ {\cal G}_{\Lambda_1}(\sigma), {\cal G}_{\Lambda_2} ({\bar \sigma}) \} = - \Big( \Lambda_1^N \partial_N \Lambda_2^M   -  \Lambda_2^N \partial_N \Lambda_1^M  \Big) \Pi_M  \delta (\sigma - {\bar \sigma}) + \kappa  \langle \Lambda_1 (\sigma), \Lambda_2 ({\bar \sigma})   \rangle  \delta^\prime (\sigma - {\bar \sigma})\, .
\end{equation}
Using (\ref{eq:fdelta}), the anomalous term can be rewritten as
\begin{eqnarray}\label{eq:}
\kappa  \langle \Lambda_1 (\sigma), \Lambda_2 ({\bar \sigma})   \rangle  \delta^\prime (\sigma - {\bar \sigma}) =
\kappa  \langle \Lambda_1 (\sigma), \Lambda_2 ({\sigma})   \rangle  \delta^\prime (\sigma - {\bar \sigma}) +
\kappa  \langle \Lambda_1 (\sigma), \Lambda_2^\prime ({\sigma})   \rangle  \delta (\sigma - {\bar \sigma}) \, ,
\end{eqnarray}
which with the help of (\ref{eq:deltapola}) can be further transformed into
\begin{eqnarray}\label{eq:}
\kappa  \langle \Lambda_1 (\sigma), \Lambda_2 ({\bar \sigma})   \rangle   \delta^\prime (\sigma - {\bar \sigma}) =
\frac{\kappa}{2}  \Big( \langle \Lambda_1, \Lambda_2^\prime    \rangle  -  \langle \Lambda_1^\prime , \Lambda_2   \rangle \Big)  \delta (\sigma - {\bar \sigma}) \nonumber \\
+  \frac{\kappa}{2}   \Big( \langle \Lambda_1, \Lambda_2  \rangle  (\sigma) + \langle \Lambda_1, \Lambda_2  \rangle  ({\bar \sigma})  \Big)
\delta^\prime (\sigma - {\bar \sigma}) \, ,
\end{eqnarray}
where the dependence of $\sigma$ has been omitted, where all terms depend solely on it.

Next, we write
\begin{eqnarray}\label{eq:}
\kappa \Lambda^{\prime M} = \kappa \, X^{\prime N} \partial_N \Lambda^{M} \, ,
\end{eqnarray}
and with the help of (\ref{eq:TD1})
\begin{eqnarray}\label{eq:}
\kappa \Lambda^{\prime M} \simeq \eta^{N R} \Pi_R \partial_N \Lambda^{M} \, .
\end{eqnarray}
The full anomalous term can now be written as
\begin{eqnarray}\label{eq:}
\kappa  \langle \Lambda_1 (\sigma), \Lambda_2 ({\bar \sigma})   \rangle   \delta^\prime (\sigma - {\bar \sigma}) =
\frac{1}{2}  \eta_{P Q} \, \eta^{M N} \Big( \Lambda_1^P \partial_N \Lambda_2^Q    - \Lambda_2^P \partial_N \Lambda_1^Q  \Big)  \Pi_M  \delta (\sigma - {\bar \sigma})   \nonumber \\
+  \frac{\kappa}{2}   \Big( \langle \Lambda_1, \Lambda_2  \rangle  (\sigma) + \langle \Lambda_1, \Lambda_2  \rangle  ({\bar \sigma})  \Big)
\delta^\prime (\sigma - {\bar \sigma})  \, .
\end{eqnarray}
The second line of the previous equation disappears after the integration with respect to $\sigma$ and $\bar{\sigma}$.

Consequently,
\begin{eqnarray}\label{eq:}
\Big\{ G_{\Lambda_1}(\sigma), G_{\Lambda_2} ({\bar \sigma}) \Big\} =&& - \Big[ \Lambda_1^N \partial_N \Lambda_2^M   -  \Lambda_2^N \partial_N \Lambda_1^M     \\ \notag
&&-
\frac{1}{2}  \eta_{P Q} \, \eta^{M N} \Big( \Lambda_1^P \partial_N \Lambda_2^Q    - \Lambda_2^P \partial_N \Lambda_1^Q  \Big)
\Big] \Pi_M  \delta (\sigma - {\bar \sigma})  \, .
\end{eqnarray}
We recognize that we can write the relation (\ref{eq:GGCbr}) as
\begin{equation} \label{eq:}
\{G(\Lambda_1), G(\Lambda_2) \} = - G \Big( [\Lambda_1, \Lambda_2]_C \Big) \, ,
\end{equation}
where $[\Lambda_1, \Lambda_2]_C $ is the $C$-bracket, given by
\begin{eqnarray}\label{eq:Cbrdef}
{[\Lambda_1, \Lambda_2]_C}^M =  \Lambda_1^N \partial_N \Lambda_2^M   -  \Lambda_2^N \partial_N \Lambda_1^M    -
\frac{1}{2}   \,  \Big( \Lambda_1^N \partial^M \Lambda_{2 N}    - \Lambda_2^N \partial^M \Lambda_{1 N}  \Big)    \, .
\end{eqnarray}
The $C$-bracket was firstly obtained in \cite{siegel1,siegel} as the generalization of the Lie derivative in the double space. For ${}^\star \pi^\mu = 0$, and  $y=0$  the double phase space reduces to the initial one, while the generator (\ref{eq:Gen}) reduces to the generator of general coordinate transformations (\ref{eq:GCTdef}), which gives rise to the Lie bracket.

We could have obtained $C$-bracket within the framework of self T-duality as well, by demanding that the parameters depend on both $x$ and $y$, substituting ${}^\star \pi^\mu = \kappa x^{\prime \mu}$ in (\ref{eq:Gen})
\begin{eqnarray}\label{eq:Gen2}
{\cal G} (\xi, \lambda ) =   \xi^\mu ( x, y )  \pi_\mu   + \kappa \lambda_\mu (x, y) x^{\prime \mu}   \,  .
\end{eqnarray}

If we additionally demand that the symmetry parameters do not depend on the T-dual coordinates $y_\mu$, this generator turns out to be exactly the Courant bracket generator (\ref{eq:GC}). It is in the accordance with \cite{doucou} that the C-bracket reduces to the Courant bracket, in case when there is no dependence on $y$.

\section{Conclusion}
\cleq
In this paper, we firstly considered the bosonic string moving in the background defined solely by the metric tensor, in which the generalized metric $G_{MN}$ has a simple diagonal form (\ref{eq:GMN}). The general coordinate transformations are generated by canonical momenta $\pi_\mu$, parametrised with vector components $\xi^\mu$. We have extended this generator, so that it is self T-dual, adding the symmetry generated by coordinate $\sigma$-derivative $ x^{\prime \mu}$, that are T-dual to the canonical momenta $\pi_\mu$ (\ref{eq:xpidual}). The extended generator of both of these symmetries is a function of a double gauge parameter $\Lambda^M$ (\ref{eq:Lxi}). The latter is a generalized vector, i.e. an element of a space obtained from a direct sum of vectors and 1-forms. The symmetry generator ${\cal G}(\Lambda) = {\cal G}(\xi,\lambda)$ of both of aforementioned symmetries was expressed as the standard $O(D,D)$ inner product of two generalized vectors (\ref{eq:gltilde}). The Poisson bracket between the extended generators ${\cal G}(\Lambda_1)$ and ${\cal G}(\Lambda_2)$ resulted up to a sign in the generator ${\cal G}(\Lambda)$, with its argument being equal to the Courant bracket of the double gauge parameters $\Lambda = [\Lambda_1, \Lambda_2]_{\cal C}$. As this is analogous to an appearance of the Lie bracket in the algebra of general coordinate transformations generators, we concluded that the Courant bracket is the self T-dual extension of the Lie bracket.

Afterwards, we added the Kalb-Ramond field $B_{\mu \nu}$ to the background, transforming the diagonal generalized metric $G_{MN}$ acting by the $B$-transformation $e^{\hat{B}}$ (\ref{eq:ebhat}). The standard generalized metric for bosonic string $H_{MN}$ was obtained (\ref{eq:genmet}), as well as the well known expressions for the Hamiltonian (\ref{eq:Hcdef}) and the Lagrangian (\ref{eq:langr}). We noted that it is possible to express the Hamiltonian in terms of the diagonal generalized metric $G_{MN}$, on the expense of transforming the double canonical variable $X^M$ by the $B$-shift. This newly obtained canonical variable $\hat{X}$ was suitable for rewriting the symmetry generator ${\cal G}$ as ${\cal G}^{\hat{B}}$, which is no longer self T-dual. This is the generator of both general coordinate, and local gauge transformations. The Poisson bracket algebra of this new generator was calculated and as an argument of the resulting generator the Courant bracket twisted by the Kalb-Ramond field was obtained. It deviates from the Courant bracket by the term related to the $H$-flux, which is the term that breaks down the self T-duality invariance. 

We considered the self T-dual description of the bosonic string $\sigma$-model. Analogously as in the first description, the complete Hamiltonian was constructed starting from the background characterized only by the T-dual metric $^\star G^{\mu \nu} = (G_E^{-1})^{\mu \nu}$. We applied the $\theta$-transformations $e^{\hat{\theta}}$ (\ref{eq:enateta}), T-dual to B-shifts, and obtained the same canonical Hamiltonian. Similarly to the previous case, the action of $\theta$-transformation on the double canonical variable was chosen for an appropriate basis. In this basis, the symmetry generator dependent upon some new gauge parameters was constructed and its algebra gave rise to the $\theta$-twisted Courant bracket. This bracket is characterized by the presence of terms related to non-geometric $Q$ and $R$ fluxes. 

It would be interesting to obtain the bracket that includes all of the fluxes, while remaining invariant upon the self T-duality. The natural candidate for this is the Courant bracket twisted by both the Kalb-Ramond field and the non-commutativity parameter. This could be done by the matrix $e^{\breve{B}}$, where 
\begin{equation}
\breve{B} = \hat{B} + \hat{\theta} = \begin{pmatrix}
0 & \kappa \theta^{\mu \nu} \\
2B_{\mu \nu} & 0 
\end{pmatrix}\, .
\end{equation}
This transformation is not trivial, as the square of the matrix $\breve{B}$ is not zero. Nevertheless, the transformation is also an element of the $O(D,D)$ group, and it remains an interesting idea for future research \cite{new}.

Lastly, we considered the symmetry generator in the double phase space that is a sum of the initial and T-dual phase space. The generator of general coordinate transformations is extended so that it includes the analogous generator in the T-dual phase space, generated by T-dual momenta ${}^\star \pi^\mu$. Both symmetry parameters were taken to depend on both the initial and T-dual coordinates, in which case the $C$-bracket is obtained as the bracket of the algebra of those generators. The $C$ bracket has already been established as the T-dual invariant bracket \cite{doucou,siegel1,siegel}, from the gauge algebra in the double space. We obtain its Poisson bracket representation, using the T-duality relations between canonical variables of different, mutually T-dual, phase spaces. These T-duality relations gave rise to the non-trivial Poisson bracket between the initial and T-dual momenta, which makes a crucial step in obtaining $C$-bracket.

We conclude that both Courant and $C$-bracket are T-dual invariant extension of the Lie bracket. The former is their extension in the initial phase space, that governs both the local gauge and general coordinate transformations. The latter is the extension of Lie bracket in the double phase space, that is a direct sum of the initial and T-dual phase space. Though the algebra of the generators that gives rise to the Courant bracket always closes, the algebra of generators in a double phase space that produces $C$-bracket only closes on a restricted parameters on an isotropic space. If all variables are independent of T-dual coordinates $y_\mu$, the $C$-bracket reduces to the Courant bracket, which confirms results from our paper.

\end{document}